\documentclass[aps,superscriptaddress]{revtex4}
\usepackage{amssymb,amsmath,epsfig}
\usepackage[colorlinks=true, pdfstartview=FitV, linkcolor=blue, citecolor=red, urlcolor=magenta, breaklinks=true]{hyperref}
\usepackage{bm}
\usepackage{amssymb}

\newcommand{\be}{\begin{eqnarray}}
\newcommand{\e}{\end{eqnarray}}
\newcommand{\bear}{\begin{eqnarray}}
\newcommand{\ear}{\end{eqnarray}}

\begin{document}

\title{Reconstructing $f(T)$ modified gravity from ECHDE and ECNADE models  }

\author{A. E. Godonou}
\email{emilog@yahoo.fr}
\affiliation{Institut de Math\'ematiques et de Sciences Physiques
(IMSP), 01 BP 613
Porto-Novo, Benin}
\author{Ines G. Salako}
\email{inessalako@gmail.com}
\affiliation{Ecole de G\'enie Rural (EGR), 01 BP 55 K\'etou, Benin}
\affiliation{Institut de Math\'ematiques et de Sciences Physiques
(IMSP), 01 BP 613
Porto-Novo, Benin}
\author{ M. J. S. Houndjo}
\email{Sthoundjo@yahoo.fr}
\affiliation{Facult\'e des Sciences et Techniques de Natitingou(FAST), BP 72, Natitingou, Benin}
\affiliation{Institut de Math\'ematiques et de Sciences Physiques
(IMSP), 01 BP 613
Porto-Novo, Benin}
 \author{Etienne Baffou}
\email{baffouh.etienne@yahoo.fr}
\affiliation{Facult\'e des Sciences et Techniques de Natitingou(FAST), BP 72, Natitingou, Benin}
\affiliation{Institut de Math\'ematiques et de Sciences Physiques
(IMSP), 01 BP 613
Porto-Novo, Benin}
\author{Joel Tossa}
\email{ines.salako@imsp-uac.org}
\affiliation{Institut de Math\'ematiques et de Sciences Physiques
(IMSP), 01 BP 613
Porto-Novo, Benin}

\begin{abstract}
We investigate alternative candidates
to dark energy that can explain the current state of the universe in the framework
of the generalized teleparallel theory of gravity $f(T)$ where $T$ denotes the torsion scalar. 
To achieve this, we carried out a
series of reconstruction taking into account to the ordinary and entropy-corrected versions of
 the holographic and new agegraphic dark energy models.
These models  used as alternative to dark energy in the literature in order  describe the current state of our universe. 
It is remarked that the models  reconstructed indicates behaviors like phantom or quintessence models. 
Furthermore, we also generated the EoS parameters associated to entropy-corrected models and we observed 
 a transition phase between quintessence state and  phantom state as showed by recent observational data. We 
also investigated on the stability theses models and we created the $\{r-s\}$.
The behavior of certains physical parameters as speed of sound and the Statefinder
parameters are compatible with current observational data.

\end{abstract}
\maketitle
\pretolerance10000
\vspace{2pc}

\section{Introduction}\label{sec1}

Some irrefutable evidence such as \cite{debamba1,debamba2,debamba3,debamba4,debamba5,debamba6} 
indicate that the universe our universe is experiencing an accelerated expansion. 
A possible candidate responsible for this current
behavior of the universe is a mysterious energy with negative
pressure which the origin and nature always  stay unelucidated. 
For reasons of incompatibility with the experimental data, several DE models proposed for this purpose have proved 
unsuccessful\cite{Padmanabhan,Copeland}). Based on the holographic principle,  \cite{Horava,Hooft,Fischler}, a promising DE candidate
so-called holographic DE (HDE) was proposed. For a good review see about HDE, see 
\cite{cohen,Li}. Interesting results have been obtained in the literature
\cite{Enqvist,Elizalde2,Guberina1,Guberina2,Karami1}.\par 
It is no longer necessary to demonstrate the importance of the black hole entropy $S_{\rm BH}$ in
the derivation of holographic dark energy. Following\cite{modak}, this  entropy-area relation $S_{\rm BH} = A/(4G)$,
where $A\sim L^2$ is the area of horizon of black hole may undergo change as 
\begin{eqnarray}
S_{\rm
BH}=\frac{A}{4G}+\tilde{\alpha}\ln{\frac{A}{4G}}+\tilde{\beta},\label{MEAR}
\end{eqnarray}
which the parameters  $\tilde{\alpha}$ and $\tilde{\beta}$ are dimensionless
constants of order unity. These modifications have been  subject of a large study in the framework loop quantum gravity (LQG) \cite{HW}. 
In this renewed attention that wei proposed the energy density of the entropy-corrected dark energy (ECHDE) model\cite{HW}. 
Besides the holographic dark energy, Cai\cite{Cai} suggested a new idea so-called 
agegraphic dark energy in order to to describe current behavior of the universe. This model has been subject of several discussions
in literature\cite{Kar1,Maz}. It should be noted that
this model can not describe the matter-dominated epoch. 
In order  to correct the insufficiency, the new agegraphic dark energy (NADE) model was proposed by Wei
$\&$ Cai \cite {Wei1} and interesting results can be found in \cite{Wei2,Kim,Kim11,Sheykhi}.
Thereafter,
Using a similar method of modification of ECHDE model, wey suggested a version corrected so-called 
entropy-corrected NADE (ECNADE) \cite{HW}
which has been investigated by \cite{Karami2}.\par
Another approach to better understand this mysterious energy is to modify standard
theories of gravity such as the General Relativity or the equivalent theory so-called Teleparallel Theory.
Some gravity theories proposed as \cite{ma1,ma2,mj1,mj5}. 
Among these proposed gravity theories, one caught our attention. This is $f(T)$ gravity as the modified version 
of the Teleparallel Theory.
Interesting  results have been found there \cite{st1}- \cite{st2}.\par
In view of $f(T)$ gravity as an alternative theory of mystery candidate,  it is necessary to investigate
 how this theory can describe the  Entropy-Corrected holographic Dark Energy (ECHDE) and 
 the Entropy-Corrected New Agegraphic Dark Energy (ECNADE) models.
 The paper is organized as follows. In section\ref{sec2}, we present the generality on $f(T)$ gravity. 
 In sections \ref{sec3} to \ref{sec6} we present a
series of reconstruction taking into account to the ordinary and entropy-corrected versions of
 the holographic and new agegraphic dark energy models. stability analysis, Statefinder parameters and the reconstruction in de Sitter space 
will be presented through the sections \ref{sec7}, \ref{sec8}, \ref{sec9} and 
the conclusion is presented in section \ref{sec10}.

\section{ Generality on $f(T)$ gravity} \label{sec2}
Unlike General Relativity and its modified versions where the gravitational interaction is described by making use
the Levi-Civita's connection, the teleparallel theory and its modified versions describe the gravitational interaction via
the curvatureless Weizenbock's connection,  whose non-null torsion defined by

\begin{eqnarray}
\Gamma^{\lambda}_{\mu\nu}=e^{\;\;\lambda}_{i}\partial_{\mu}e^{i}_{\;\;\nu}=-e^{i}_{\;\;\mu}\partial_\nu e_{i}^{\;\;\lambda}.
\end{eqnarray}
Thus, we can define respectively torsion and the contorsion, the torsion scalar as
\begin{eqnarray}
T^{\lambda}_{\;\;\;\mu\nu}= \Gamma^{\lambda}_{\mu\nu}-\Gamma^{\lambda}_{\nu\mu},
\end{eqnarray}
\begin{eqnarray}
K^{\mu\nu}_{\;\;\;\;\lambda}=-\frac{1}{2}\left(T^{\mu\nu}_{\;\;\;
\lambda}-T^{\nu\mu}_{\;\;\;\;\lambda}+T^{\;\;\;\nu\mu}_{\lambda}\right)\,\,,
\end{eqnarray}
\begin{eqnarray}
T=T^{\lambda}_{\;\;\;\mu\nu}S^{\;\;\;\mu\nu}_{\lambda}\,.
\end{eqnarray}
Now, we define the action of the modified version of TG whose Lagrangian is an algebraic function depending of the torsion.
\begin{eqnarray}
 S= \int e \left[\frac{f(T)}{2\kappa^2} +\mathcal{L}_{m} \right]d^{4}x   \label{eq9}\,,
\end{eqnarray}
where $\kappa^{2} = 8 \pi G $ is the usual gravitational
coupling constant. The following equations of motion are  obtained by varying the the action (\ref{eq9}) via the tetrads
\begin{eqnarray}
S^{\;\;\; \nu \rho}_{\mu} \partial_{\rho} T f_{TT} + 
[e^{-1} e^{i}_{\;\; \mu}\partial_{\rho}(e e^{\;\; \mu}_{i}S^{\;\;\; \nu\lambda}_{\alpha} )
+T^{\alpha}_{\;\;\; \lambda \mu}   S^{\;\;\; \nu \lambda}_{\alpha} ]f_{T}+
\frac{1}{4}\delta^{\nu}_{\mu}f=\frac{\kappa^{2}}{2} \mathcal{T}^{\nu}_{\mu}  \label{eq10}\,,
\end{eqnarray}
where $\mathcal{T}^{\nu}_{\mu}$ is the energy 
momentum tensor, $f_{T} = df(T)/dT$ and 
$f_{TT}  = d^{2}f(T)/dT^{2}$ the first and second derivative of $f(T)$ with respect to $T$. 
Considering a
flat $FLRW$ cosmologies, the torsion scalar  yields to
\begin{eqnarray}
 T= -6H^{2}, \label{R}
\end{eqnarray}
where $H=\dot{a}/a$ denotes the Hubble parameter. 
Thus, the usual Friedmann equations become

\begin{eqnarray}
 H^2 &=& \frac{8\pi G}{3}(\rho+\rho_T), \label{FiEq1}\\
 2\dot H+3H^2 &=& -\frac{8\pi G}{3}(p+p_T), \label{FiEq2}
\end{eqnarray}

where
\begin{eqnarray}
 \rho_T &=& \frac{1}{16\pi G}[2T f_{T}-f-T], \label{density} \\
 p_T &=& \frac{1}{16\pi G}[2\dot H(4T f_{TT}+2 f_{T}-1)]-\rho_T. \label{pT}
\end{eqnarray}
This above quantities define respectively the  torsion contributions to the energy density and pressure 
who checks the following continuity equation.
 \begin{eqnarray}
  \dot{\rho}_{T}+3H(\rho_{T}+p_{T}) &=& 0.\label{ecT}
 \end{eqnarray}
We  can notice for  the null torsion contribution,  the general relativity (Teleparallel Gravity)  recoved. 
In order to preserve the modified nature of the gravity and  by make using Eqs. (\ref{density}) and (\ref{pT}), we can write
the effective torsion equation of state as \cite{Nozari}
\begin{eqnarray}\label{wHDETotal}
    \omega_{T}\equiv\frac{p_T}{\rho_T}=-1-\frac{4\dot H \Big(2Tf''(T)+f'(T)-1 \Big)}{T -2Tf'(T) + f(T) }.
\end{eqnarray}

Considering a universe for a slow contribution of the matter, we can note that the Eq. (\ref{FiEq1}) yields
\begin{eqnarray}
\frac{3}{k^2}H^2=\rho_T.
\end{eqnarray}
Thus, the EoS parameter  can be obtained in the same conditions as previously by making use (\ref{wHDETotal}) and (\ref{ecT})
\begin{eqnarray}
\omega_{T}=-1-\frac{2\dot{H}}{3H^2}.
\end{eqnarray}
We can remark that the first derivative of the Hubble parameter is positive $\dot{H}>0$ 
for an accelerating expanded phantom-like
universe ($\omega_{T}<-1$) and negative $\dot{H}<0$ for an accelerated expanded quintessence-like one $\omega_{T}>-1$.
Based on the torsion contribution and any class of scale factor $a=a(t)$, one  can perform
a series of reconstruction taking according to the ordinary and entropy-corrected versions of
 the holographic and new agegraphic dark energy models. Among the existing series of classes of scale factors, we will focus 
on two classes habitually used in modified gravity to describe the process of the accelerating
universe \cite{Nojiri}.

The first category of scale factor is defined by \cite{Nojiri,Sadjadi}
\begin{eqnarray}
a(t) = a_0\,(t_s-t)^{-h}, \quad t\leq t_s, \quad h>0.\label{a}
\end{eqnarray}
By considering  Eqs. (\ref{R}) and (\ref{a}), the Hubble parameter becomes
\begin{eqnarray}
H=\frac{h}{t_s-t}=\left[\frac{-T}{6} \right]^{1/2},~~~\dot{H}=H^2/h,\label{respect
to r}
\end{eqnarray}
which shows that 

The model (\ref{a}) the model responds perfectly to a model that describes an accelerating expanded phantom-like
universe $\dot{H}=H^2/h>0$ and this is also the reason why the said model is  habitually  so-called the phantom-like
scale factor.

On the other hand, we can define the second category of scale factor  as \cite{Nojiri}
\begin{eqnarray}
a(t)=a_0t^h,~~~h>0,\label{aQ}
\end{eqnarray}
and yields to 
\begin{eqnarray}
H=\frac{h}{t}=\left[\frac{-T}{6}  \right]^{1/2},~~~\dot{H}=-H^2/h,\label{respect
to rQ}
\end{eqnarray}
which $\dot{H} <0$. We can remark  that the model (\ref{aQ}) indicate 
a accelerating expanded quintessence-like universe.
In order to give a  cosmological interpretation, we plotted   versus redshift $z=\frac{a_0}{a} -1$. Thus, for the 
 first category of scale factor (\ref{a}) become
\begin{eqnarray}\label{first}
 T= \frac{-6 h^2}{(1+z)^{\frac{2}{h}}},
\end{eqnarray}
on the other hand the second category of scale factor (\ref{aQ}) yields
 
 \begin{eqnarray}\label{second}
 T= -6 h^2(1+z)^{\frac{2}{h}}.
\end{eqnarray}
In the following sections,  
we present a series of reconstruction taking into account to the ordinary and entropy-corrected versions of
 the holographic and new agegraphic dark energy models and  by making use of of two categories of scale factors 
 previously  described.
 
\section{$f(T)$ reconstruction from HDE model}\label{sec3}
Following\cite{Li}, the holographic dark energy model in a spatially flat
universe is characterized by a density   given by 
\begin{eqnarray}
\rho_{\Lambda}=\frac{3c^2}{k^2R_h^2},\label{ro H}
\end{eqnarray}
where $c$ is a numerical constant introduced for convenience.
The latest observational data for a flat universe
show that this constant  worth
 $c=0.818_{-0.097}^{+0.113}$ \cite {Li6}. We define radius of the event horizon  $R_h$ as 
\begin{eqnarray}
R_h=a\int_t^{\infty}\frac{{\rm d}t}{a}=a\int_a^{\infty}\frac{{\rm
d}a}{Ha^2}.\label{L0}
\end{eqnarray}
Thus, we can evaluate for first category  of scale factor (\ref{a}) the future event horizon $R_h$ by making use  Eq.
(\ref{respect to r})
\begin{eqnarray}
R_h=a\int_t^{t_s}\frac{{\rm
d}t}{a}=\frac{t_s-t}{h+1}=\frac{h}{h+1}\sqrt{\frac{-6}{T}}.\label{L}
\end{eqnarray}
Substituting Eq. (\ref{L}) into (\ref{ro H}) one can obtain
\begin{eqnarray}
\rho_\Lambda=\frac{- 3 c^2 (h+1)^2  }{6 k^2 h^2}T.\label{ro H R}  
\end{eqnarray}
Replacing  Eq. (\ref{ro H R}) in the differential equation
(\ref{density}), i.e. $\rho_T=\rho_{\Lambda}$, yields the following
solution
\begin{eqnarray}
f(T)=\frac{\left(h^2-c^2 (1+h)^2\right) T}{h^2}+\sqrt{T} C_1 ,\label{frHDE}
\end{eqnarray}
where $C_1$ is integration constant. 
For more consistency, the constants have been to be determined and to to do so, we impose 
the initial conditions\cite{}, assuming the assumption according to what, at present time, 
the holographic model must recover the usual $\Lambda$CDM  one, that is
\begin{eqnarray}
\left(f\right)_{t=t_0}=T_0-2\Lambda\;\;,\quad\quad \left(\frac{df}{dt}\right)_{t=t_0}=\left(\frac{dT}{dt}\right)_{t=t_0}\;,\label{icond}
\end{eqnarray}
where the subscript $t_0$ and $T_0$ denote the present time and the related value of the torsion scalar, respectively.\par
Making use of the initial conditions (\ref{icond}), one gets
\begin{eqnarray}
C_1= \frac{-2 h^2 \Lambda +c^2 (1+h)^2 T_0}{h^2 \sqrt{T_0}}  
\end{eqnarray}
such that the algebraic  holographic dark energy model reads
\begin{eqnarray}
f(T)=\frac{\left(h^2-c^2 (1+h)^2\right) T}{h^2}+\frac{\sqrt{T}
\left(-2 h^2 \Lambda +c^2 (1+h)^2 T_0\right)}{h^2 \sqrt{T_0}}\;.\label{genemodel}
\end{eqnarray}

We can obtain easily the EoS parameter for $f(T)$-gravity  according to the HDE  model by 
replacing Eq. (\ref{genemodel}) into (\ref{wHDETotal}) and making use 
(\ref{respect to r}), 
\begin{eqnarray} 
\omega_{T}=-1-\frac{2}{3h},~~~h>0,\label{wHDE2}
\end{eqnarray}
which data indicates  to a accelerating expanded phantom-like universe, i.e.
$\omega_T<-1$. We see also  that the EoS
parameter $\omega_{T}$  according to the observational data \cite{Copeland}.

Considering the second class of scale factor (\ref{aQ}) and 
(\ref{respect to rQ}),  the future event horizon $R_h$ yields to
\begin{eqnarray}
R_h=a\int_t^{\infty}\frac{{\rm
d}t}{a}=\frac{t}{h-1}=\frac{h}{(h-1)}\sqrt{\frac{-6}{T}},~~~h>1. \label{RhQ}
\end{eqnarray}
Also, using the same procedure, we can determine
\begin{eqnarray}
 f(T) =\frac{\left(-c^2 (-1+h)^2+h^2\right) T}{h^2}+\frac{\sqrt{T} \left(-2 h^2 \Lambda +c^2 (-1+h)^2 T_0\right)}{h^2 \sqrt{T_0}},
\end{eqnarray}
and 
\begin{eqnarray}
\omega_{T}=-1+\frac{2}{3h},~~~h>1.\label{wHDEQ}
\end{eqnarray}
We can note that the EoS parameter satisfies $-1<\omega_T<-1/3$ which data indicates  to 
a accelerating expanded quintessence-like universe.
%-------------------------------------------------------------------------------------
\section{$f(T)$ reconstruction from ECHDE model}\label{sec4}
In this section, we present the modified version of energy density suggested  by Wei \cite{HW}, so-called (ECHDE). This model is
given by
\begin{eqnarray}
\rho_\Lambda=\frac{3c^2}{k^2R_h^2}+\frac{\alpha}{R_h^4}\ln\left(\frac{R_h^2}{k^2}\right)+\frac{\beta}{R_h^4},\label{ECHDE}
\end{eqnarray}
with two dimensionless constants $\alpha$ and $\beta$.
unity. One note that the HDE model is recoved  for $\alpha=\beta=0$.  Thus, the last two
terms in Eq. (\ref{ECHDE}) constitute the  corrections made to the HDE model. When the universe becomes large, ECHDE
reduces to the ordinary HDE \cite{HW}.

Considering the first category of scale factor (\ref{a}) and making use the Eq.(\ref{L}), (\ref{ECHDE}), 
the density corresponding to entropy-corrected HDE becomes
\begin{eqnarray}
\rho_\Lambda=\frac{(1+h)^2 T \left((1+h)^2 T \beta -\frac{18 c^2 h^2}{\kappa ^2}+(1+h)^2 T \alpha 
\text{Log}\left[-\frac{6 h^2}{(1+h)^2 T \kappa ^2}\right]\right)}{36
h^4}.\label{ECHDE
R}
\end{eqnarray}
Thus, the algebraic function  according to ECHDE model yields
\begin{eqnarray}
f(T)&=&\sqrt{T} C_1 + \frac{T \left(-162 h^2 (c+(-1+c) h) (c+h+c h) \right) }{162 h^4}    
\\ &+& \;\; \frac{ T \left( (1+h)^4 T (2 \alpha +3 \beta ) \kappa ^2     +3 (1+h)^4T \alpha  \kappa ^2
\text{Log}\left[-\frac{6 h^2}{(1+h)^2 T \kappa ^2}\right]\right)}{162 h^4},    \label{frECHDE}
\end{eqnarray}
where $C_1$ is constant. By making use the boundary conditions (\ref{icond}), we can determine $C_1$ as
\begin{eqnarray}
C_1= -\frac{2 \Lambda }{\sqrt{T_0}}+\frac{c^2 (1+h)^2 \sqrt{T_0}}{h^2}-\frac{(1+h)^4 \kappa ^2 \left(2 \alpha +3 \beta +3
\alpha  \text{Log}\left[-\frac{6 h^2}{(1+h)^2 \kappa ^2 T_0}\right]\right) T_0^{3/2}}{162 h^4},\label{c1}
\end{eqnarray}
which yields the algebraic function $f(T)$ model  according to the  entropy-corrected HDE.
\begin{eqnarray}
f(T)&=&\frac{T \left(-162 h^2 (c+(-1+c) h) (c+h+c h)+(1+h)^4 T (2 \alpha +3 \beta ) \kappa ^2 \right)}{162 h^4}  \cr
&+&    \frac{ T\left( 3 (1+h)^4 T 
\alpha  \kappa ^2 \text{Log}\left[-\frac{6
h^2}{(1+h)^2 T \kappa ^2}\right]\right)}{162 h^4}  +  \sqrt{T} \left(-\frac{2 \Lambda }{\sqrt{T_0}}+
\frac{c^2 (1+h)^2 \sqrt{T_0}}{h^2} \right)      \cr
  & -&  \sqrt{T} \left( \frac{(1+h)^4
\kappa ^2 \left(2 \alpha +3 \beta +3 \alpha  \text{Log}\left[-\frac{6 h^2}{(1+h)^2 
\kappa ^2 T_0}\right]\right) T_0^{3/2}}{162 h^4}\right).\label{c2}
\end{eqnarray}
Replacing Eq. (\ref{frECHDE}) into (\ref{wHDETotal}) we determine
the EoS parameter corresponding to the algebraic function according to ECHDE model
\begin{eqnarray}
\omega_{T}  &=& -1 -   
    \Big\{ \frac{4 \left(18 c^2 h^2+(1+h)^2 T (\alpha -2 \beta ) \kappa ^2 \right)}{T \left(-18 c^2 h^2+(1+h)^2 T \beta  \kappa ^2+(1+h)^2 T \alpha  \kappa ^2 \text{Log}\left[-\frac{6 h^2}{(1+h)^2
T \kappa ^2}\right]\right)} \\
    &-& \frac{ 8 \left( (1+h)^2 T \alpha  \kappa ^2 \text{Log}\left[-\frac{6 h^2}{(1+h)^2 T
\kappa ^2}\right]\right) H'[t]}{T \left(-18 c^2 h^2+(1+h)^2 T \beta  \kappa ^2+(1+h)^2 T \alpha  \kappa ^2 \text{Log}\left[-\frac{6 h^2}{(1+h)^2
T \kappa ^2}\right]\right)} \Big\}, 
~~~h>0, \;\;\;\;  \label{wHDE3}
\end{eqnarray}
which can be rewritten under a form similar to the relation (\ref{wHDE2}) 
\begin{eqnarray}
\omega_{T} &=& -1 
 - \frac{2}{3\,h}\Bigg\{ \frac{ \left(3 c^2 h^2-(1+h)^2 H^2 (\alpha -2 \beta )
 \kappa ^2 \right)}{\left(3 c^2 h^2+(1+h)^2 H^2 \beta  \kappa ^2+(1+h)^2 H^2 \alpha
\kappa ^2 \text{Log}\left[\frac{h^2}{(1+h)^2
H^2 \kappa ^2}\right]\right)}   \\
 &+& \frac{ \left( 2 (1+h)^2 H^2 \alpha
\kappa ^2 \text{Log}\left[\frac{h^2}{(1+h)^2 H^2
\kappa ^2}\right]\right) }{ \left(3 c^2 h^2+(1+h)^2 H^2 \beta  \kappa ^2+(1+h)^2 H^2 \alpha
\kappa ^2 \text{Log}\left[\frac{h^2}{(1+h)^2
H^2 \kappa ^2}\right]\right)}\Bigg\},~~~h>0.\;\;\;\;\;\; \label{wECHDE}
\end{eqnarray}
We can observe that the figure\ref{fig1} presents presents a phase transition between   
  quintessence state i.e $\omega_T>-1$, and 
the phantom state i.e $\omega_T<-1$.

Considering the second category of scale factor (\ref{aQ}), we can determine the density corresponding 
to  the entropy-corrected holographic dark energy model as
\begin{eqnarray}
 \rho_\Lambda =  \frac{(-1+h)^2 T \left((-1+h)^2 T \beta -\frac{18 c^2 h^2}{\kappa ^2}+
 (-1+h)^2 T \alpha  \text{Log}\left[-\frac{6 h^2}{(-1+h)^2 T \kappa
^2}\right]\right)}{36 h^4}. \;\;\;\;
\end{eqnarray}
Using the above procedure, the algebraic function corresponding to entropy-corrected holographic dark energy model yields

\begin{eqnarray}
 f(T) &=&  \frac{T \left(-162 c^2 (-1+h)^2 h^2+162 h^4+(-1+h)^4 T (2 \alpha +3 \beta ) \kappa ^2 \right)}{162 h^4} \cr
  &+&   \frac{ T \left( 3 (-1+h)^4 T \alpha  \kappa ^2 \text{Log}\left[-\frac{6
h^2}{(-1+h)^2 T \kappa ^2}\right]\right)}{162 h^4} 
+ \sqrt{T} \left(-\frac{\Lambda }{\sqrt{T_0}}+\frac{c^2 (-1+h)^2 \sqrt{T_0}}{h^2} \right) \cr
&-& \sqrt{T} \left( \frac{(-1+h)^4
\kappa ^2 \left(2 \alpha +3 \beta +3 \alpha  \text{Log}\left[-\frac{6 h^2}{(-1+h)^2 \kappa ^2 T_0}\right]\right) T_0^{3/2}}{162 h^4}\right)
\end{eqnarray}

where we making use the boundary conditions (\ref{icond}) for determine the integration constant.
Now, we can easily determine for this second category of scale factor,
the EoS parameter corresponding to the algebraic function according to entropy-corrected holographic dark energy model
$f(T)$-gravity model 
\begin{eqnarray}
\omega_{T} &=& -1 - \Big\{ 
\frac{4 \left(18 c^2 h^2+(-1+h)^2 T (\alpha -2 \beta ) \kappa ^2 \right)}{T \left(-18 c^2 h^2+(-1+h)^2 T \beta  \kappa ^2+(-1+h)^2 T \alpha  \kappa ^2 \text{Log}\left[-\frac{6 h^2}{(-1+h)^2
T \kappa ^2}\right]\right)} \cr
  &-&  \frac{ 8 \left(  (-1+h)^2 T \alpha  \kappa ^2 \text{Log}\left[-\frac{6 h^2}{(-1+h)^2
T \kappa ^2}\right]\right) H'[t]}{T \left(-18 c^2 h^2+(-1+h)^2 T \beta  \kappa ^2+(-1+h)^2 T \alpha  \kappa ^2 \text{Log}\left[-\frac{6 h^2}{(-1+h)^2
T \kappa ^2}\right]\right)}  \Big \}
\end{eqnarray}
which can be rewritten under a form similar to the relation (\ref{wHDEQ}) 
\begin{eqnarray}
\omega_{T} &=& -1 + \frac{2}{3\,h} \Big\{- \frac{ \left(3 c^2 h^2-(-1+h)^2 H^2 (\alpha -2 \beta ) \kappa ^2 \right)}{\left(3 c^2 h^2+(-1+h)^2 H^2 \beta  \kappa ^2+(-1+h)^2 H^2 \alpha  \kappa ^2 \text{Log}\left[\frac{h^2}{(-1+h)^2
H^2 \kappa ^2}\right]\right)} \cr
  &+& \frac{ \left( 2 (-1+h)^2 H^2 \alpha  \kappa ^2 \text{Log}\left[\frac{h^2}{(-1+h)^2
H^2 \kappa ^2}\right]\right) H'[t]}{ \left(3 c^2 h^2+(-1+h)^2 H^2 \beta  \kappa ^2+(-1+h)^2 H^2 \alpha  \kappa ^2 \text{Log}\left[\frac{h^2}{(-1+h)^2
H^2 \kappa ^2}\right]\right)}  \Big \}
\end{eqnarray}
As observed previously in the case of the first class of the scaling factor, we note through the figure\ref{fig1} 
a phase transition between $\omega_T>-1$
and $\omega_T<-1$.
%%%%%%%%%%%%%%%%%%%%%%%%%%%%%%%%%%%%%%%%%%%%%%%%%%%%%%%%%%%%%%%%%%%%%%%%%%%%%%%%%%%%%%%%%%%%%%%
%-------------------------------------------------------------------------------------
\section{$f(T)$ reconstruction from NADE model}\label{sec5}

 the Agegraphic dark Energy(ADE) model suggested by Cai  describe the actuel state of the universe \cite{Cai} but
 based\cite{Kar1,Maz}. In order to explain the
matter-dominated epoch, the  New Agegraphic dark Energy ADE (NADE) model was proposed by Wei
$\&$ Cai \cite {Wei1}, while the time scale was chosen to be the
conformal time instead of the age of the universe.
The energy density and the conformal time  corresponding respectively to this model are given by
\begin{eqnarray}
\rho_{\Lambda}=\frac{3{n}^2}{k^2\eta^2},\label{NADE}
\end{eqnarray}
and
\begin{eqnarray}
\eta=\int\frac{{\rm d}t}{a}=\int\frac{{\rm d}a}{Ha^2}.\label{eta}
\end{eqnarray}
%%%%%%%%%%%%%%%%%%%%%%%%%%%%%%%%%%%%%%%%%%%%%%%%%%%%%%%%%%%%%%%%%%%%%%%%%%%%%%%%
Considering the first category of scale factor (\ref{a}), and making use the Eq. (\ref{respect to r}),  the conformal time
$\eta$ yields
\begin{eqnarray}
\eta=\int_t^{t_s}\frac{{\rm
d}t}{a}=\frac{(t_s-t)^{h+1}}{a_0(h+1)}=\frac{h^{h+1}}{a_0(h+1)}\left(\frac{-6}{T}\right)^{\frac{h+1}{2}}.\label{E}
\end{eqnarray}
We can determine the density corresponding by replacing Eq. (\ref{E}) into (\ref{NADE})
\begin{eqnarray}
\rho_\Lambda=-\frac{2^{-1-h} 3^{-h} h^{-2 (1+h)} (1+h)^2 n^2 \left(-\frac{1}{T}\right)^{-h} T a_0^2}{\kappa ^2},\label{ro
NEDE R}
\end{eqnarray}
which yields to the differential equation where the solution is given by
\begin{eqnarray}
f(T)=T+\sqrt{T} C_1-\frac{6^{-h} h^{-2 (1+h)} (1+h)^2 n^2 \left(-\frac{1}{T}\right)^{-h} T a_0^2}{1+2 h}.\label{frNADE}
\end{eqnarray}
  Using the boundary
conditions (\ref{icond}), we can obtain the integration constant as
\begin{eqnarray}
C_1=\frac{-2\Lambda +\frac{6^{-h} h^{-2 (1+h)} (1+h)^2 n^2 a_0^2 \left(-\frac{1}{T_0}\right){}^{-h} T_0}{1+2 h}}{\sqrt{T_0}} ,
\end{eqnarray}
Thus, we can determine  $f(T)$-gravity model  according to  the new Agegraphic dark Energy (NADE) model as
\begin{eqnarray}
f(T)&=&T+\frac{6^{-h} h^{-2 (1+h)} (1+h)^2 n^2 \left(-\frac{1}{T}\right)^{-h} \sqrt{T} a_0^2 \left(-\sqrt{T} 
\left(-\frac{1}{T_0}\right){}^h+\left(-\frac{1}{T}\right)^h
\sqrt{T_0}\right) \left(-\frac{1}{T_0}\right){}^{-h}}{1+2 h} \cr
&-&\frac{2\,\sqrt{T} \Lambda }{\sqrt{T_0}}.
\end{eqnarray}
Now, we can easily determine for this first category of scale factor,
the EoS  corresponding to the algebraic function according to the new Agegraphic dark Energy (NADE).
\begin{eqnarray}
 \omega_{T}=-1+\frac{4 (1+h) H'[t]}{T}.
\end{eqnarray}

Substituting Eq. (\ref{frNADE}) into (\ref{wHDETotal}), we get 
EoS  corresponding to the algebraic function according to the new Agegraphic dark Energy (NADE)
\begin{eqnarray}
\omega_{T}=-1-\frac{2(h+1)}{3h},~~~h>0,\label{wNADE}
\end{eqnarray}
We can note that the EoS parameter satisfies $\omega_T<-1$ which data indicates  to
a accelerating expanded quintessence-like universe.

Considering the second category of scale factor (\ref{aQ}) and using (\ref{respect
to rQ}) we obtain 
\begin{eqnarray}
\eta=\int_0^t\frac{{\rm
d}t}{a}=\frac{t^{1-h}}{a_0(1-h)}=\frac{h^{1-h}}{a_0(1-h)}\left(\frac{-6}{T}\right)^{\frac{1-h}{2}}.
~~~\frac{1}{2}<h<1,\label{EQ}
\end{eqnarray}
In order to preserve a real finite conformal time, it is necessary to to have
the condition $\frac{1}{2}<h<1$. 
Replacing Eq. (\ref{E}) into (\ref{NADE}) one can get the density
\begin{eqnarray}
\rho_\Lambda=-\frac{2^{-1+h} 3^h (-1+h)^2 h^{-2+2 h} n^2 \left(-\frac{1}{T}\right)^h T a_0^2}{\kappa ^2},\label{ro
NEDE R1}
\end{eqnarray}
and the algebraic function as
\begin{eqnarray}
f(T)=T+\sqrt{T} C_1+\frac{6^h (-1+h)^2 h^{-2+2 h} n^2 \left(-\frac{1}{T}\right)^h T a_0^2}{-1+2 h},\label{frNADE1}
\end{eqnarray}
where $C_1$ is integration constant.  Using the boundary
conditions (\ref{icond}), the constant and the algebraic function
$f(T)$ are determined as
\begin{eqnarray}
C_1=\frac{-2\Lambda +\frac{6^h (-1+h)^2 h^{-2+2 h} n^2 a_0^2 \left(-\frac{1}{T_0}\right){}^{-1+h}}{-1+2 h}}{\sqrt{T_0}},
\end{eqnarray}
\begin{eqnarray}
f(T)=T+\frac{6^h (-1+h)^2 h^{-2+2 h} n^2 \sqrt{T} a_0^2 \left(\left(-\frac{1}{T}\right)^h \sqrt{T}-
\left(-\frac{1}{T_0}\right){}^h \sqrt{T_0}\right)}{-1+2
h}-\frac{2\sqrt{T} \Lambda }{\sqrt{T_0}}.
\end{eqnarray}
Thus, we can deduce the EoS parameter as

\begin{eqnarray}
 \omega_{T}=-1-\frac{4 (-1+h) H'[t]}{T},
\end{eqnarray}
which can be rewritten using (\ref{respect
to rQ})
\begin{eqnarray}
\omega_{T}=-1+\frac{2(1-h)}{3h},~~~\frac{1}{2}<h<1.\label{wNADEQ}
\end{eqnarray}
We can note that the EoS parameter satisfies $-1<\omega_T<-1/3$ which data indicates  to
a accelerating expanded quintessence-like universe.
%-------------------------------------------------------------------------------------
\section{$f(T)$ reconstruction from ECNADE model}\label{sec6}
Based on a similar method to the ECHDE model, the entropy-corrected 
NADE was suggested by Wei \cite{HW} and later  by \cite{Karami2} which the energy density is defined by 
\begin{eqnarray}
\rho_{\Lambda} = 
\frac{\alpha}{\eta^4}\ln{\left(\frac{\eta^2}{k^2}\right)} +
\frac{\beta}{\eta^4} + \frac{3n^2}{k^2\eta^2}.\label{ECNADE}
\end{eqnarray}

Considering the first category of scale factor (\ref{a}) and replacing Eq.
(\ref{E}) into (\ref{ECNADE}), the density corresponding becomes
\begin{eqnarray}
\rho_\Lambda & =&\frac{1}{\kappa ^2}36^{-1-h} h^{-4 (1+h)} (1+h)^2 \left(-\frac{1}{T}\right)^{-2 h} T a_0^2 \Big\{-2^{1+h} 3^{2+h} c^2 h^{2+2 h} 
\left(-\frac{1}{T}\right)^h \cr
 &+&(1+h)^2
T \kappa ^2 \left(\beta +\alpha  \text{Log}\left[\frac{6^{1+h} h^{2+2 h} \left(-\frac{1}{T}\right)^{1+h}}{(1+h)^2
\kappa ^2 a_0^2}\right]\right)
a_0^2\Big\}.\label{ro
ECNEDE R}
\end{eqnarray}
Thus, the entropy-corrected agegraphic dark energy $f(T)$-gravity model yields

\begin{eqnarray}
f(T)&=&T+\sqrt{T}C_1-\frac{6^{-h} c^2 h^{-2 (1+h)} (1+h)^2 \left(-\frac{1}{T}\right)^{-h} T a_0^2}{1+2 h}\cr
&+&\frac{2^{-1-2 h} 9^{-1-h} h^{-4 (1+h)} (1+h)^4 \left(-\frac{1}{T}\right)^{-2 h} T^2 \kappa ^2 \Big\{2 (1+h) \alpha +(3+4 h) \beta }{(3+4 h)^2} \cr
  &+& \frac{ (3+4 h) \alpha
 \text{Log}\left[\frac{6^{1+h} h^{2+2 h} \left(-\frac{1}{T}\right)^{1+h}}{(1+h)^2 \kappa ^2 a_0^2}\right]\Big\} a_0^4}{(3+4 h)^2},~~
\label{frECNADE}
\end{eqnarray}
where $C_1$ is integration constant which can be  determined from
the boundary conditions (\ref{icond}) as
\begin{eqnarray}
C_1 &=& -\frac{\Lambda}{\sqrt{T_0}} 
  -  \frac{2^{-1-2 h} 9^{-1-h} h^{-4 (1+h)} (1+h)^4 \kappa ^2 \Big\{ 2 (1+h) \alpha }{\sqrt{T_0} (3+4 h)^2} \cr
  &+& \frac{ (3+4 h) \beta +(3+4 h)
\alpha  \text{Log}\left[\frac{6^{1+h} h^{2+2 h} \left(-\frac{1}{T_0}\right){}^{1+h}}{(1+h)^2 \kappa ^2 a_0^2}\right]\Big\} a_0^4 
\left(-\frac{1}{T_0}\right){}^{-2
(1+h)}}{ \sqrt{T_0} (3+4 h)^2} \cr
   &+&  \frac{6^{-h} c^2 h^{-2 (1+h)} (1+h)^2 a_0^2 \left(-\frac{1}{T_0}\right){}^{-h} T_0}{ \sqrt{T_0} (1+2 h)}.\label{c3}
\end{eqnarray}
Therefore, the algebraic  entropy-corrected agegraphic dark energy $f(T)$-gravity model reads
\begin{eqnarray}
f(T) &=& T+\frac{6^{-h} c^2 h^{-2 (1+h)} (1+h)^2 \left(-\frac{1}{T}\right)^{-h} \sqrt{T} a_0^2 \Big\{-\sqrt{T} \left(-\frac{1}{T_0}\right){}^h}{1+2 h} \cr
   &+ & \frac{ \left(-\frac{1}{T}\right)^h
\sqrt{T_0}\Big\} \left(-\frac{1}{T_0}\right){}^{-h}}{1+2 h} 
  - \frac{\sqrt{T} \Lambda }{\sqrt{T_0}} 
  + \Big\{ \frac{1}{(3+4 h)^2}2^{-1-2 h} 9^{-1-h} h^{-4 (1+h)}
(1+h)^4 \Big\} \cr
  & \times & \Big\{ \left(-\frac{1}{T}\right)^{-2 h} \sqrt{T} \kappa ^2 a_0^4 \left(-\frac{1}{T_0}\right)^{-2 h} \Big\} \cr
  & \times & \Bigg\{(T^{3/2} \left(2 (1+h) \alpha +(3+4 h)
\beta +(3+4 h) \alpha  \text{Log}\left[\frac{6^{1+h} h^{2+2 h} \left(-\frac{1}{T}\right)^{1+h}}{(1+h)^2 \kappa ^2 a_0^2}\right]\right) 
\left(-\frac{1}{T_0}\right){}^{2
h} \cr
   &-& \left(-\frac{1}{T}\right)^{2 h} \Big\{ 2 (1+h) \alpha +(3+4 h) \beta \cr
   &+& (3+4 h) \alpha  \text{Log}\left[\frac{6^{1+h} h^{2+2 h}
\left(-\frac{1}{T_0}\right){}^{1+h}}{(1+h)^2
\kappa ^2 a_0^2}\right]\Big\} T_0^{3/2}\Bigg\}.\label{c4}
\end{eqnarray}
Substituting Eq. (\ref{frECNADE}) into (\ref{wHDETotal}), the EoS
parameter corresponding to algebraic function according to ECNADE
model

\begin{eqnarray}
\omega_{T}=-\left(2^{1+h} 3^{2+h} c^2 h^{2+2 h} \left(-\frac{1}{T}\right)^h \left(-T+4 (1+h) H'[t]\right)+(1+h)^2 T \kappa ^2 a_0^2 
\left(T \left(\beta
+\alpha  \text{Log}\left[\frac{6^{1+h} h^{2+2 h} \left(-\frac{1}{T}\right)^{1+h}}{(1+h)^2 \kappa ^2 a_0^2}\right]\right)
-4 (1+h) \left(-\alpha +2
\beta +2 \alpha  \text{Log}\left[\frac{6^{1+h} h^{2+2 h} \left(-\frac{1}{T}\right)^{1+h}}{(1+h)^2 \kappa ^2 a_0^2}\right]\right)
H'[t]\right)\right)/\left(T
\left(-2^{1+h} 3^{2+h} c^2 h^{2+2 h} \left(-\frac{1}{T}\right)^h  
+(1+h)^2 T \kappa ^2 \left(\beta +\alpha 
\text{Log}\left[\frac{6^{1+h} h^{2+2 h}
\left(-\frac{1}{T}\right)^{1+h}}{(1+h)^2 \kappa ^2 a_0^2}\right]\right) a_0^2\right)\right).\label{wECNADE1}
 \end{eqnarray}

\begin{figure}[h]
\begin{minipage}{14pc}
\includegraphics[width=16pc]{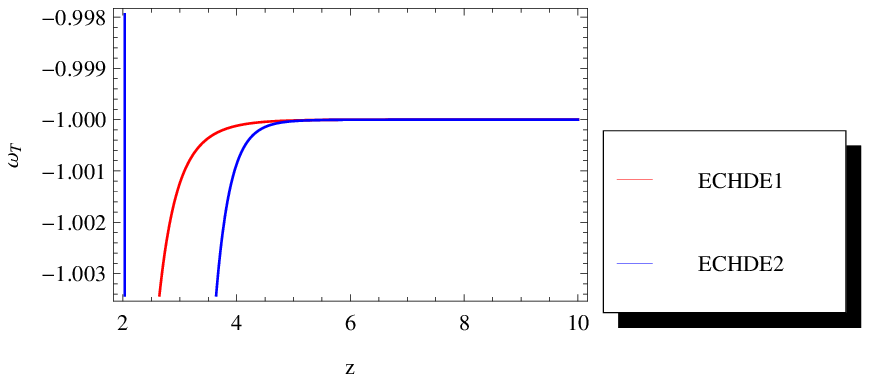}
\caption{\label{fig1}Plot of $\omega_{T}$ versus $z$. The curves of ECHDE model 
are characterize as red is for the first category (\ref{first}), blue
is for the second category (\ref{second}).}
\end{minipage}\hspace{3pc}%
\begin{minipage}{14pc}
\includegraphics[width=16pc]{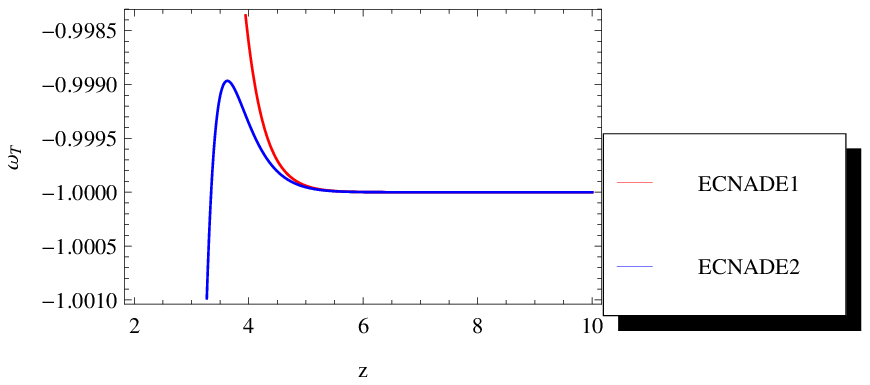}
\caption{\label{fig2}Plot of $\omega_{T}$ versus $z$. The curves of ECNADE model 
are characterize as red is for the first category (\ref{first}), blue
is for the second category (\ref{second}).}
\end{minipage}\hspace{3pc}%
\end{figure}

Though the figure\ref{fig2} we can remark a phase transition between $\omega_T>-1$ and 
$\omega_T<-1$. 

Considering the second category of scale factor (\ref{aQ}), introducing   Eq.
(\ref{E}) into (\ref{ECNADE}) one obtains
\begin{eqnarray}
\rho_\Lambda &=& 6^{-2+h} (-1+h)^2 h^{-4+2 h} \left(-\frac{1}{T}\right)^h T a_0^2 \Bigg\{-\frac{18 c^2 h^2}{\kappa ^2} 
+6^h (-1+h)^2 h^{2 h} \cr
 & \times & \left(-\frac{1}{T}\right)^h
T \left(\beta +\alpha  \text{Log}\left[\frac{6^{1-h} h^{2-2 h} \left(-\frac{1}{T}\right)^{1-h}}{(-1+h)^2 
\kappa ^2 a_0^2}\right]\right) a_0^2\Bigg\}.\label{ro
ECNEDE R2}
\end{eqnarray}
Solving the differential equation (\ref{density}) for the energy
density (\ref{ro ECNEDE R}) yields
\begin{eqnarray}
f(T)&=& T+\sqrt{T} C_1+\frac{6^h c^2 (-1+h)^2 h^{-2+2 h} \left(-\frac{1}{T}\right)^h T a_0^2}{-1+2 h}+\cr
&+& 
\frac{2^{-1+2 h} 9^{-1+h} (-1+h)^4 h^{-4+4 h} \left(-\frac{1}{T}\right)^{-1+2 h} T \kappa ^2 \Bigg\{ 2 (-1+h) \alpha}{(3-4 h)^2} \cr
 &+& \frac{ (-3+4 h) \beta +(-3+4 h)
\alpha  \text{Log}\left[\frac{6^{1-h} h^{2-2 h} \left(-\frac{1}{T}\right)^{1-h}}{(-1+h)^2 \kappa ^2 a_0^2}\right]\Bigg\} a_0^4}{(3-4 h)^2},~~
\label{frECNADE2}
\end{eqnarray}
where $C_1$ is integration constant. Also $C_1$ is determined from
the boundary conditions (\ref{icond}) as
\begin{eqnarray}
C_1 &=&\frac{-\Lambda +\frac{6^h c^2 (-1+h)^2 h^{-2+2 h} a_0^2 \left(-\frac{1}{T_0}\right){}^{-1+h}}{-1+2 h} }{\sqrt{T_0}} \cr
  & +&  \frac{2^{-1+2 h}
9^{-1+h} (-1+h)^4 h^{-4+4 h} \kappa ^2 \Bigg\{2 (-1+h) \alpha +(-3+4 h) \beta }{ \sqrt{T_0} (3-4 h)^2} \cr
&+& \frac{ (-3+4 h) \alpha  \text{Log}\left[\frac{6^{1-h} h^{2-2 h} 
\left(-\frac{1}{T_0}\right){}^{1-h}}{(-1+h)^2
\kappa ^2 a_0^2}\right]\Bigg\} a_0^4 \left(-\frac{1}{T_0}\right){}^{2 h} T_0^2}{ \sqrt{T_0} (3-4 h)^2},\label{c32}
\end{eqnarray}
Therefore, the algebraic function for the second class of scale factor (\ref{aQ})  according to ECNADE model reads
\begin{eqnarray}
f(T)&=&T+\frac{6^h c^2 (-1+h)^2 h^{-2+2 h} \sqrt{T} a_0^2 \left(\left(-\frac{1}{T}\right)^h \sqrt{T}-\left(-\frac{1}{T_0}\right){}^h \sqrt{T_0}\right)}{-1+2h}
  \cr  &-& \frac{\sqrt{T} \Lambda }{\sqrt{T_0}}-\frac{1}{(3-4 h)^2}2^{-1+2 h} 9^{-1+h} (-1+h)^4 h^{-4+4 h} \sqrt{T} \kappa ^2 a_0^4 \cr
 & \times & \Bigg\{ \left(-
\frac{1}{T}\right)^{2
h} T^{3/2} \Big\{ 2 (-1+h) \alpha +(-3+4 h) \beta \cr
  &+& (-3+4 h) \alpha  \text{Log}\left[\frac{6^{1-h} h^{2-2 h} \left(-\frac{1}{T}\right)^{1-h}}{(-1+h)^2
\kappa ^2 a_0^2}\right]\Big\} +\Big\{ 2 \alpha -2 h \alpha +3 \beta -4 h \beta \cr
  &+& (3-4 h) \alpha  \text{Log}\left[\frac{6^{1-h} h^{2-2 h} \left(
-\frac{1}{T_0}\right){}^{1-h}}{(-1+h)^2
\kappa ^2 a_0^2}\right]\Big\} \left(-\frac{1}{T_0}\right){}^{2 h} T_0^{3/2}\Bigg\}.\label{c42}
\end{eqnarray}
introducing Eq. (\ref{frECNADE}) into (\ref{wHDETotal}) yields the EoS
parameter corresponding to the algebraic function according to  the ECNADE  
model as
\begin{eqnarray}
\omega_{T}=\left(18 c^2 h^2 \left(T+4 (-1+h) H'[t]\right)-6^h (-1+h)^2 h^{2 h} \left(-\frac{1}{T}\right)^h T \kappa ^2 a_0^2 
\left(T \left(\beta
+\alpha  \text{Log}\left[\frac{6^{1-h} h^{2-2 h} \left(-\frac{1}{T}\right)^{1-h}}{(-1+h)^2 \kappa ^2 a_0^2}\right]\right) 
+ 4 (-1+h) \left(-\alpha
+2 \beta +2 \alpha  \text{Log}\left[\frac{6^{1-h} h^{2-2 h} \left(-\frac{1}{T}\right)^{1-h}}{(-1+h)^2 
\kappa ^2 a_0^2}\right]\right) H'[t]\right)\right)/\left(T
\left(-18 c^2 h^2+6^h (-1+h)^2 h^{2 h} \left(-\frac{1}{T}\right)^h  T \kappa ^2 \left(\beta +\alpha 
\text{Log}\left[\frac{6^{1-h} h^{2-2 h} \left(-\frac{1}{T}\right)^{1-h}}{(-1+h)^2
\kappa ^2 a_0^2}\right]\right) a_0^2\right)\right). \label{wECNADE12}
\end{eqnarray}
Here, we observe through the figure\ref{fig2} the same behavior as previously.

\section{Analysis of reconstructed model} \label{sec7}
In order to analysis the reconstructed models, we  check in the section 
the behavior of certains physical parameters as speed of sound and the Statefinder parameters.
Squared speed of sound, 
\begin{eqnarray}\label{sound}
v_s^2=\frac{\dot{p}_{eff}}{\dot{\rho}_{eff}},
\end{eqnarray}
is an important quantity to test the stability of the background
evolution. 
The study of the stability of the model will depend on the sign of (\ref{sound}).
Several discussions were conducted leading to interesting results \cite{setare3,kim,jawad,Chattopadhyay}. 
 We here considered the $v_s^2$ as equal to (\ref{sound}
and plot this $v_s^2$ verses the cosmic time $z$ for both reconstructions of $f(T)$ ECHDE and ECNADE models.

\begin{figure}[h]
\begin{minipage}{14pc}
\includegraphics[width=16pc]{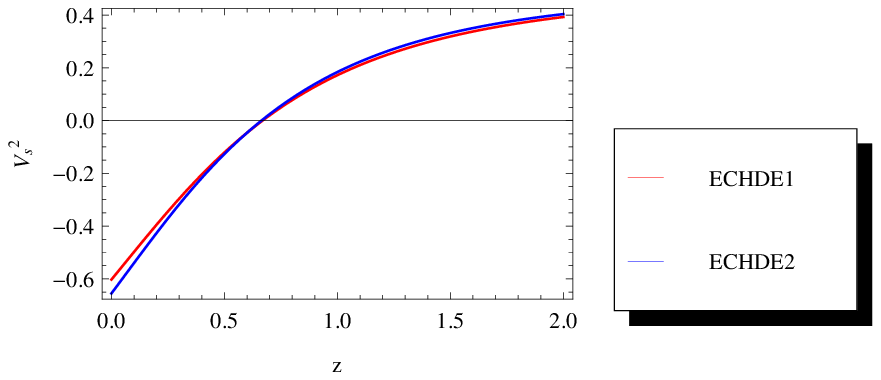}
\caption{\label{fig3}Plot of $v_s^2$ versus $z$. The curves of ECHDE model 
are characterize as red is for the first category (\ref{first}), blue
is for the second category (\ref{second}).}
\end{minipage}\hspace{3pc}%
\begin{minipage}{14pc}
\includegraphics[width=16pc]{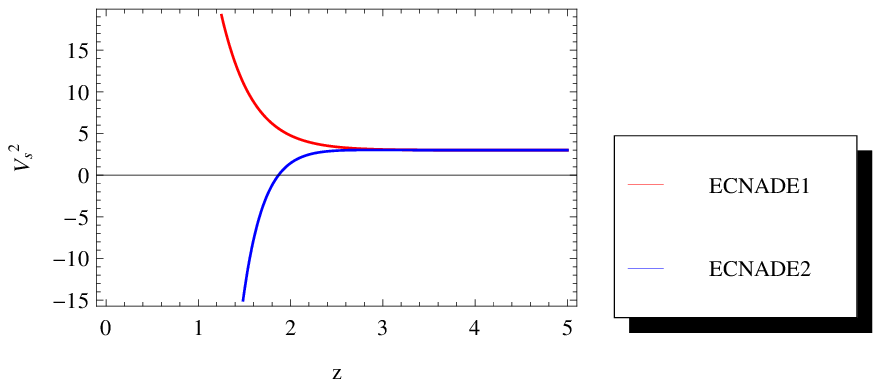}
\caption{\label{fig4}Plot of $v_s^2$ versus $z$. The curves of ECNADE model 
are characterize as red is for the first category (\ref{first}), blue
is for the second category (\ref{second}).}
\end{minipage}\hspace{3pc}%
\end{figure}
From the figures\ref{fig3}, we can observe the instability of ECHDE model for the two  categories of scale factor 
$0<z<0.6$ but the  ECHDE model is stable when $z>0.6$. The figure\ref{fig4} indicates that the 
second category of scale factor corresponding to ECNADE model  is unstable for $0<z<2$ and stable when $z>2$.
On the other hand, for the first category corresponding to the ECNADE model remains stable whatever the value of $z$.

\section{Statefinder parameters}\label{sec8}
\begin{figure}[h]
\begin{minipage}{14pc}
\includegraphics[width=16pc]{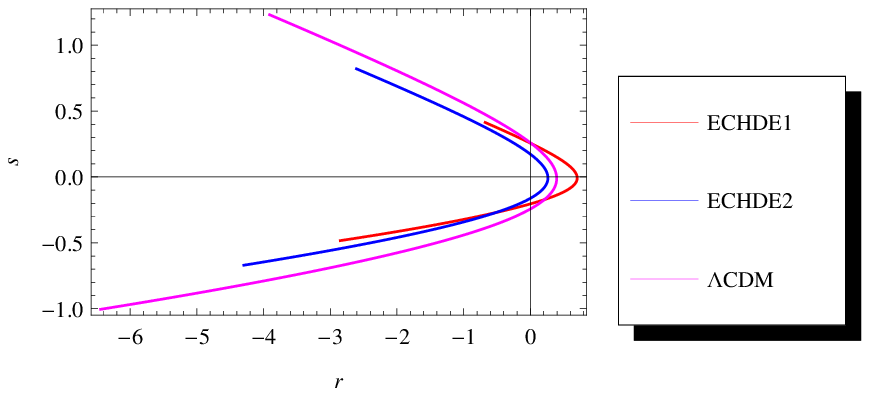}
\caption{\label{fig5}Plot of $\{r-s\}$. The curves of ECHDE model 
are characterize as red is for the first category (\ref{first}), blue
is for the second category (\ref{second}).}
\end{minipage}\hspace{3pc}%
\begin{minipage}{14pc}
\includegraphics[width=16pc]{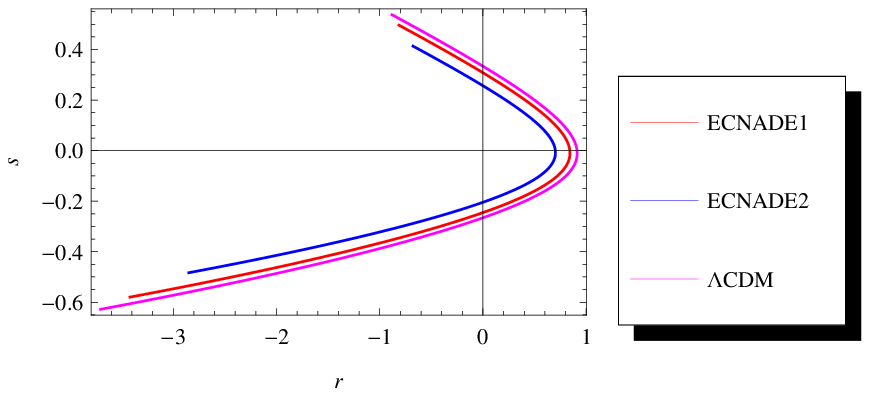}
\caption{\label{fig6}Plot of $\{r-s\}$. The curves of ECNADE model 
are characterize as red is for the first category (\ref{first}), blue
is for the second category (\ref{second}).}
\end{minipage}\hspace{3pc}%
\end{figure}

A series of candidates for the Dark Energy model have been suggested till date and it is 
noted between them some problems of competitivity. In order to
try to solve these problems, Sahni et al. 
\cite{varun1} introduced the statefinder
$\{r,s\}$ diagnostic pair. This pair  reads
\cite{varun1,varun2}
\begin{eqnarray}\label{16}
r=\frac{\dddot{a}}{H^{3}\,a}, \quad
s=\frac{r-1}{3\,\left(q-\frac{1}{2}\right)},
\end{eqnarray}
with $q$ and $H$, respectively the deceleration and  Hubble parameters. 
Thus, we can determine geometrically  properties of Dark Energy. Interesting 
 results have been found there \cite{sharif,Panotopoulos,Chakraborty,varun1,varun2,Zimdahl}.
We  created the $\{r-s\}$
trajectories and compared with the $\Lambda$CDM limit based on the reconstructed models. We can observe from the figures\ref{fig5}, \ref{fig6}  that
although the trajectories approache the $\Lambda$CDM limit. Finally, the behavior of  physical parameters as speed of sound and the Statefinder
parameters are compatible with current observational data.

 \section{Conclusions}\label{sec10}

 In this paper, we  investigated how the theory of modified gravity $ f(T) $ where $T$ denotes
the torsion scalar can describe  the  Entropy-Corrected holographic Dark Energy (ECHDE) and 
 the Entropy-Corrected New Agegraphic Dark Energy (ECNADE) models.
To achieve this, We have proceded to reconstruct the different algebraic function according to the HDE, ECHDE, NADE 
and ECNADE models by making use of of two categories of scale factors  previously  described. 
The EoS parameters have been determined  and  we note that for
 the first class of scale factor, the EoS parameters
 indicates  to a accelerating expanded phantom-like universe, i.e.
$\omega_T<-1$. We see from figures\ref{fig1},\ref{fig2} also  that the EoS parameter $\omega_{T}$  according to the current observational data. On the other hand, for the second class
of scale factor, We can note from figures\ref{fig1},\ref{fig2} that the EoS parameter satisfies $-1<\omega_T<-1/3$ 
which data indicates  to 
a accelerating expanded quintessence-like universe. 
To ensure viability of the reconstructed models, we performed a stability analysis in
In order to ensure viability of the reconstructed models, we performed a stability analysis by  checking 
the behavior of certains physical parameters as speed of sound and the Statefinder parameters.
 From the figures\ref{fig3}, we can observe the instability of ECHDE model for the two  class of scale factor 
$0<z<0.6$ but the  ECHDE model is stable when $z>0.6$. The figure\ref{fig4} indicates that the 
second category of scale factor from ECNADE model  is unstable for $0<z<2$ and stable when $z>2$. 
On the other hand, for the first category of scale factor, the ECNADE model remains stable whatever the value of $z$. 
Finally, we have created the $\{r-s\}$ trajectories and compared with the $\Lambda$CDM limit. 
We can observe from the figures\ref{fig5}, \ref{fig6}  that although the trajectories approache the $\Lambda$CDM limit.
Finally, the behavior of  physical parameters as speed of sound and the Statefinder
parameters are compatible with current observational data.

\end{document}